\documentclass[12pt]{spieman}  
\usepackage{amsmath,amsfonts,amssymb}
\usepackage{graphicx}
\usepackage{setspace}
\usepackage[margin=1in, top=1in, bottom = 1in]{geometry}
\usepackage{tocloft}
\usepackage{lipsum}
\usepackage{lineno}
\usepackage{soul}

\title{Primary Objective Grating Telescopy: Optical Properties and Feasibility of Applications}

\author[a]{Leaf Swordy}
\author[a]{Heidi Jo Newberg}
\author[a,b]{Thomas Ditto}
\affil[a]{Rensselaer Polytechnic Institute, Heidi Jo Newberg, Dept. of Physics, Applied Physics, and Astronomy, 110 8th St, Troy, NY 12180}
\affil[b]{3DeWitt LLC, Thomas D. Ditto, PO Box 10, Ancramdale, NY 12503}

\cftpagenumbersoff{figure}
\cftpagenumbersoff{table} 
\begin{document} 
\maketitle

\begin{abstract}
We develop the theoretical foundation for primary objective grating (POG) telescopy. In recent years, a wide range of telescope designs that collect the light over a large grating and focus it with a secondary receiving optic that is placed at grazing exodus have been proposed by Thomas D. Ditto, and are sometimes referred to as Dittoscopes. Applications include discovery and characterization of exoplanets, discovery of near-Earth asteroids, and spectroscopic surveys of the sky. These telescopes would have small aerial mass, and therefore provide a path forward to launch large telescopes into space. Because this series of telescope designs departs from traditional telescope designs, it has been difficult to evaluate which applications are most advantageous for this design. Here, we define a new figure of merit, the ``modified \'etendue," that characterizes the photon collection capability of a POG. It is demonstrated that the diffraction limit for observations is determined by the length of the grating. We evaluate the effects of atmospheric seeing for ground-based applications and the disambiguation of position vs. wavelength in the focal plane using a second dispersing element. Finally, some strategies for fully reaping the benefits of POG optical characteristics are discussed.
\end{abstract}


\section{Introduction}
\label{sect:intro}  

The concept of large scale space telescopes consisting of thin-film optical holograms has garnered interest since the advent of optical holography in the 1960s\cite{Baez}. The typical application is that of a large diameter Fresnel Zone Plate (FZP) or photon sieve, with proposed applications seeking to solve problems associated with UV optics and large space apertures with low aerial mass. Development of these technologies stagnated after the 1960s, likely due to promising advances in UV/IR optics coupled with the highly chromatic behavior of holographic primaries. More recently, interest in large-aperture thin-film holographic primaries has undergone a resurgence, following the development of the dispersion-corrected Fresnel lens. \cite{patent,10.1117/12.7977006}

Throughout the late 1990s and early 2000s, many different thin-film space observatories have been proposed\cite{Eye,Moire,Andersen:07,refId0} and explored to various degrees of development. Such designs feature extremely large apertures with low aerial mass and less susceptibility to typical figure errors than conventional telescopes. The primary drawback with many of the proposed concepts is found in the extremely long focal lengths they incorporate. Long focal lengths are needed in order to correct for chromatism and achieve low surface figure requirements, this topic is discussed in Hyde EYEGLASS\cite{Eye}. These focal lengths are often on the scale of kilometers, requiring precise formation flying of many optical components. 

The topic of this publication differs from the preceding discussion in that, rather than an FZP or photon sieve, the primary optical element is imagined as a large area diffraction grating at grazing exodus, subsequently viewed by a secondary telescope or camera. Placing a grating in front of the aperture of a telescope is not a new concept in and of itself; prior to the advent of optical fiber spectroscopy such techniques, generally referred to as slitless spectroscopy,  were commonplace\cite{Wesse}. The focal plane images obtained with slitless spectroscopy are similar to the more modern technique of using a grism to obtain so called `grism images'. In a grism image, each object in the focal plane is spread out into a small spectrum, thereby allowing for rudimentary spectral analysis of many objects simultaneously. The use of a diffraction grating as a primary objective dates to the very beginning of 20th century astronomy\cite{Hertz}. The spectra obtained with such techniques are generally low resolution, with higher resolution limited by spreading and obfuscation of objects in the focal plane. Thus, the advent of astronomical fiber spectroscopy in the late 1970s\cite{Med} has resulted in fewer applications utilizing primary objective grating spectroscopy. The optical designs discussed here differ from a traditional grism image due to the high exodus angle at which light is collected from the grating. This subtle change results in a wide array of interesting (and sometimes counterintuitive) optical properties.

Inspiration for this publication is drawn from more recent realizations by Thomas D. Ditto. Beginning in the early 2000s,\cite{Ditto03}  Ditto formulated unique telescope architectures featuring dispersive primaries. These eponymous `Dittoscope' architectures are unique from all previous applications, in that they utilize the POG for optical leverage by collecting light from the grating in a grazing exodus configuration, and use a secondary spectrograph to disambiguate the overlapping spectra from an enlarged field-of-view. Viewing the POG at grazing exodus can result in a massive increase in collection area and angular resolution, and a hugely dispersed spectrum of each object. It will be seen in the details of this publication that a wealth of interesting optical properties arise from the most basic realization of a Dittoscope.

Ditto has proposed many specific designs and applications of POG technology\cite{Ditto07}. Current design concepts feature a variety of light collection schemes after diffraction from the grating, with most featuring secondary spectrographic capability to disambiguate dispersed light of focal plane objects. This technique is called `dual dispersion'. While consideration of specific applications will be briefly discussed, they are generally outside the scope of this paper. 

Presented here is a purely theoretical study of the Dittoscope concept. While many real-life implementation issues exist, they will only be discussed to a very limited extent in Section \ref{sect:issue}. We intend to explore only the most basic optical properties of a POG in grazing exodus configuration, often modeling the secondary focusing element as a mere aperture-mirror combination of little description. It is the goal of this publication to place the general properties of such an arrangement on a firm foundation. The optical geometry of arbitrary systems and configurations are discussed with the goal of establishing relationships between the light collection area, field of view, bandwidth, and angular resolution. New relationships and figures of merit are described, facilitating comparison between theoretical POG architectures and contemporary observational techniques. 

We show that grazing exodus POG configurations present many potential benefits, including extremely large light collection area, high spectral/angular resolution, a massively enhanced field of view, and the potential for low-cost lightweight deployment in space. These serve as the primary motivating factors for this work. The feasibility and potential benefits of grazing exodus POG observatories are discussed.

\section{THE MOST}

\begin{figure}
\begin{center}
\begin{tabular}{c}
\includegraphics[height=6cm]{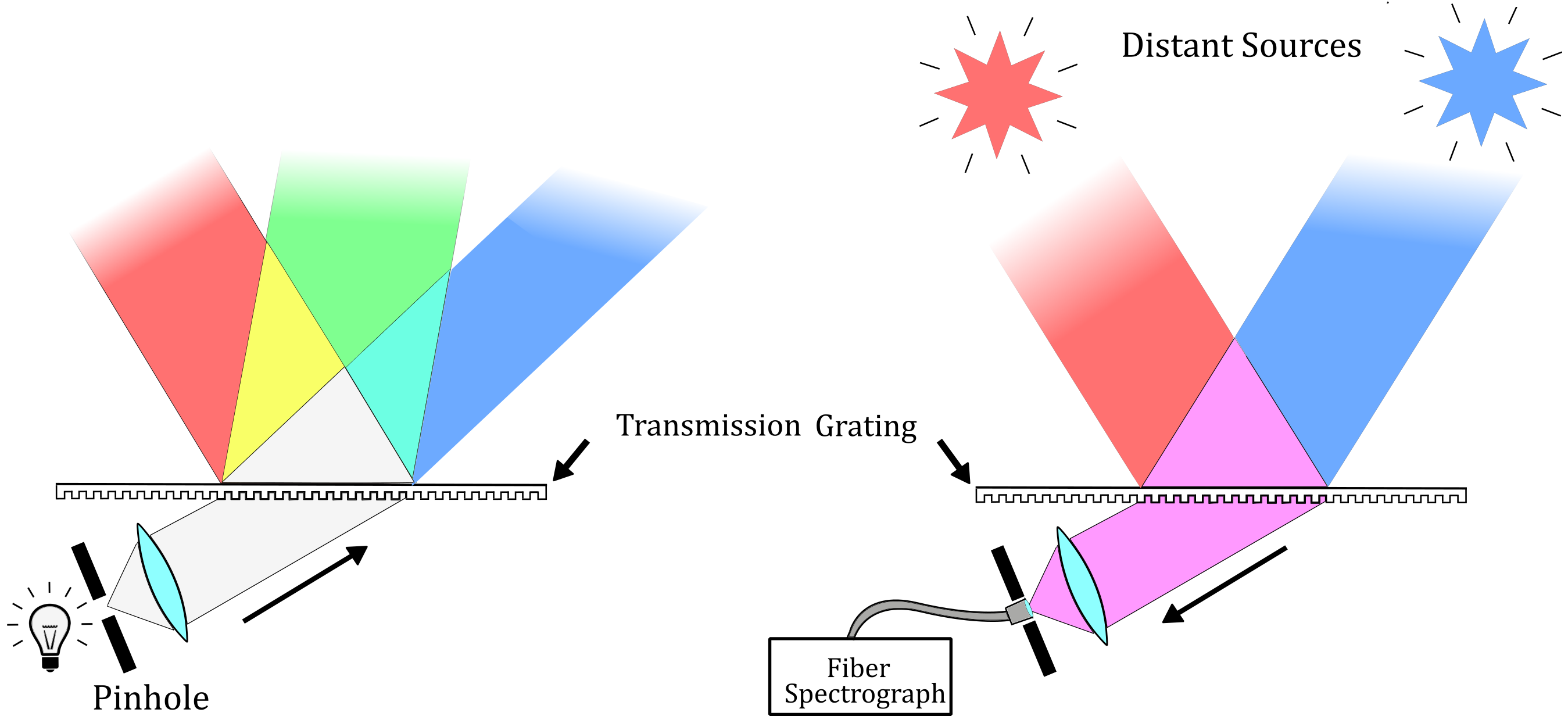}
\\
(a) \hspace{6cm} (b)
\end{tabular}
\end{center}
\caption 
{ \label{fig:trace}
(a) A broadband source placed behind a pinhole in the focal plane is output as a variety of collimated monochromatic beams. (b) The situation of (a) is reversed, such that collimated input beams from distant sources (e.g. stars in the sky) are received by an optical fiber placed in the focal plane of a receiving lens. Each location in the sky can be received only in a narrow bandwidth, allowing on-sky position to be identified by received wavelength via a spectrograph. 
}
\end{figure} 

As an introduction to the basic concepts underlying a Dittoscope, we can imagine a polychromatic collimated beam incident on a transmission grating (Figure \ref{fig:trace}a). In this scenario, the polychromatic incident beam is output as a spectrum of monochromatic exodus beams, with exodus angles determined by the grating equation:

\begin{equation}
\label{eq:gratingeq}
\textrm{sin}(\theta_{in}) + \textrm{sin}(\theta_{out}) = m\frac{\lambda}{p} \;  ,
\end{equation}

\noindent where $\theta_{in}$ can be either the angle of incidence of the polychromatic beam (Figure \ref{fig:trace}a), or the incident angle from a single object in the sky (Figure \ref{fig:trace}b), $\theta_{out}$ is the angle at which light with wavelength $\lambda$ is diffracted from the grating, $m$ is the diffraction order, and $p$ is the grating pitch. 

The two situations shown in Figure \ref{fig:trace} are essentially opposites of one another. In Figure \ref{fig:trace}a, light radiated from a pinhole in the focal plane is output to the field as a spectrum of collimated beams. Conversely, Figure \ref{fig:trace}b places a single optical fiber in the focal plane, this optical fiber receives field photons at a different wavelength according to the incident angle of the light (location on the sky). The optical arrangement of Figure \ref{fig:trace}b embodies the most basic aspects of the Dittoscope concept. The arrangement will later be complicated by moving the receiving optics to an extreme angle, and allowing the entire focal plane to receive incident photons (i.e. many fibers/slits).

\begin{figure}
\begin{center}
\begin{tabular}{c}
\includegraphics[height=8cm]{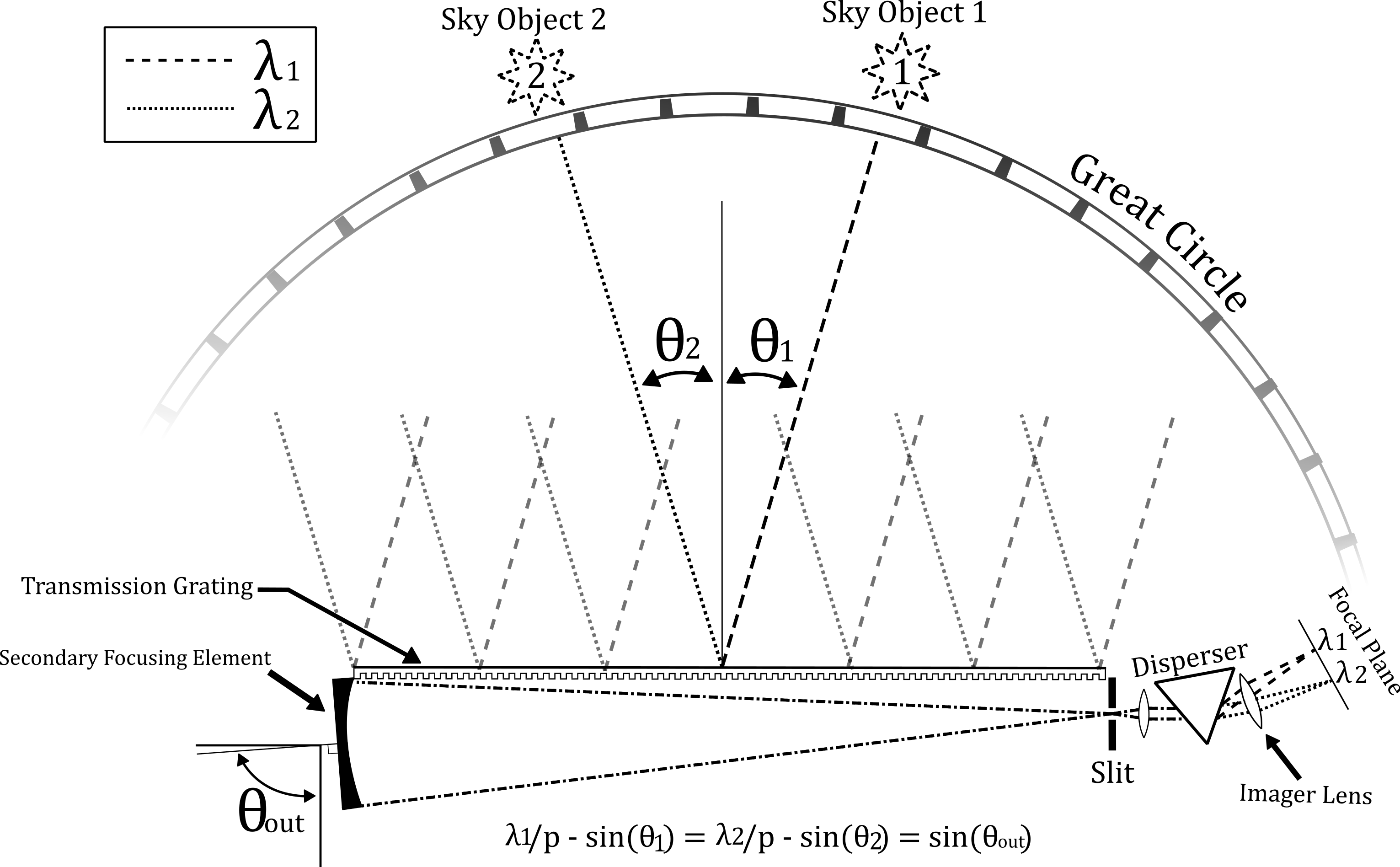}
\end{tabular}
\end{center}
\caption 
{ \label{fig:pog}
Configuration of The High \'Etendue Multiple Object Spectrographic Telescope (THE MOST). Incident angles $\boldsymbol{\theta}_{\textbf{\textit{1, 2}}}$ are diffracted into a fixed exodus angle $\boldsymbol{\theta}_{\textbf{\textit{out}}}$ and collected by the paraboloid mirror labeled `secondary focusing element'. The secondary focusing element coveys light to a slit located in the focal plane, after which the on-sky position of the incident light is disambiguated by a secondary dispersing element (conceptually represented as a prism). In this arrangement, each unique location in the sky is associated with a unique position in the focal plane, and is received in a narrow bandwidth (determined by the resolvance of the primary grating and secondary disperser). 
}
\end{figure} 

Now we will discuss a specific Dittoscope design, The High \'Etendue Multiple Object Spectrographic Telescope (THE MOST)\cite{Ditto14,MOST}. THE MOST design (Figure~\ref{fig:pog}) imagines collecting light from the sky through a large (tens of meters or larger), precision ruled primary objective plane grating. The light from this grating is then focused with a paraboloid mirror, or possibly a conventional telescope with a wider corrected field of view (in either case termed a `secondary focusing element'). In the focal plane of the secondary focusing element, the positions of objects in the sky are correlated with the wavelength of the light that is received in the direction of dispersion according to the grating equation (Equation \ref{eq:gratingeq}). In the direction perpendicular to the dispersion (i.e. oblique incident angles), the angular position of the object observed is unambiguous.

A functional difference between the classical objective prism design and THE MOST is the replacement of the prism with a grating at grazing exodus. This change allows us to image a very large collecting area from a large grating, paired with a much smaller focusing optic. The disambiguation of received photons by a secondary dispersing element (dual dispersion) is also a unique feature. In most cases, the secondary focusing element collects light at an exodus angle approaching $\theta_{out}\rightarrow90^\circ$. This secondary focusing optic could have the optical design of an ordinary optical telescope, which focuses light over a larger field of view than a simple paraboloid mirror. By collecting the light from a large exodus angle, we can ``observe" a long grating with a telescope mirror of modest size (due to anamorphic magnification). THE MOST uses the second dispersing element, or spectrograph, to untangle the position in the sky from which each photon originated from the wavelength of the photon that was received. If the position and wavelength were known for all photons received at the focal plane of the secondary, true angular positions of objects in the sky can be found by simple application of the grating equation (Eqn. ~\ref{eq:gratingeq}).

The placement of the secondary focusing element at a large angle of diffraction leads to a very high dispersion of the light from the grating. This means that only a small wavelength range of light from a particular object in the sky is spread across the entire field of view of the secondary focusing element. For an object with an angle of incidence of zero and an angle of diffraction near $90^\circ$ in the first diffraction order, the resolvance is $R = N \approx L/\lambda$ (since $\lambda \rightarrow p$ as the exodus angle approaches $90^\circ$), where N is the total number of grooves, $L$ is the length of the grating and $\lambda$ is the wavelength of the light. For a grating that is 100m long and a detection wavelength of 500nm, $R \approx 200,000,000$. Clearly, the accuracy with which wavelengths can be determined will be limited by real-life limitations such as optical defects of the grating (groove spacing and regularity), the allowable number of pixels in the focal plane, and the resolvance of the secondary disperser. However, this example provides an idea of how much a Dittoscope spreads out the light from each object.

\section{POG Telescope Geometry}

Here we explore the optical geometry of POG telescope designs in all generality, with the ultimate goal of determining a useful figure of merit by which to judge POG designs against conventional telescopes. 

\subsection{Light Collection Area}
\label{sect:area1}

\begin{figure}
\begin{center}
\begin{tabular}{c}
\includegraphics[height=5.6cm]{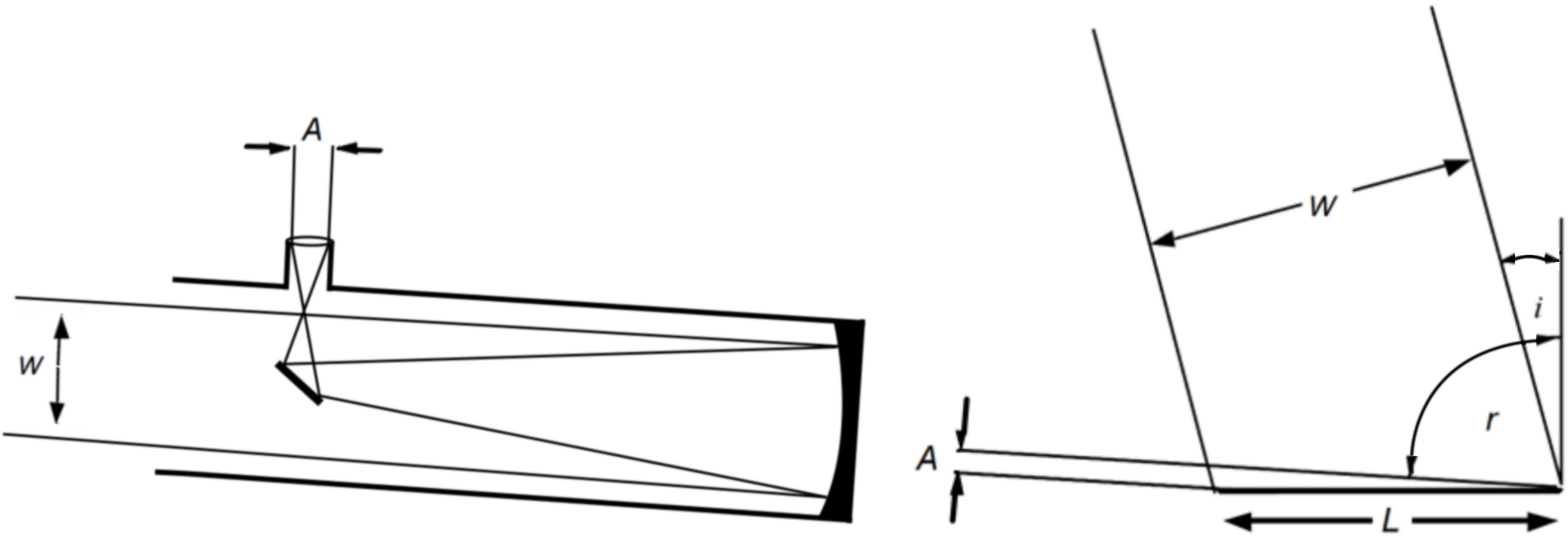}
\end{tabular}
\end{center}
\caption 
{ \label{fig:mag}
(Left) Conventional Newtonian telescope, conveying input light at width `W' to the ocular at width `A'. (Right) Anamorphic magnification from a reflection grating, monochromatic light entering at width `W' is output at width `A' after diffraction from the grating. The very large cross sectional area at W compared to the received beam A, acts in a similar manner to a conventional telescope in its ability to concentrate light.} 
\end{figure}

One of the initial selling points for the Dittoscope is the large collecting area of the low aerial mass grating, which requires a much smaller focusing element due to anamorphic magnification. The right panel of Figure~\ref{fig:mag} demonstrates that a monochromatic plane wave collected by the entire grating at width $W$ is reflected into a grazing exodus angle, resulting in a smaller cross-sectional area of width $A$ at the output. For broadband sources, each wavelength is collected at a different angle within the FOV of the secondary collector. We will postpone such discussions for the time being, instead concentrating on the single wavelength view of a grating as a telescope analog.

 In a sense, a telescope can be thought of as a device that conveys a large input beam at the primary to a narrow output beam at the ocular. An ocular-equipped telescope also has the important property of enhancing angular differences between input sources (conservation of \'etendue), so that the diffraction limit incurred by the human pupil may be overcome when making astronomical observations. In Section~\ref{sect:res} it will be shown that the Dittoscope, in the single-wavelength view,  possesses this property as well.

\begin{figure}
\begin{center}
\begin{tabular}{c}
\includegraphics[height=5.5cm]{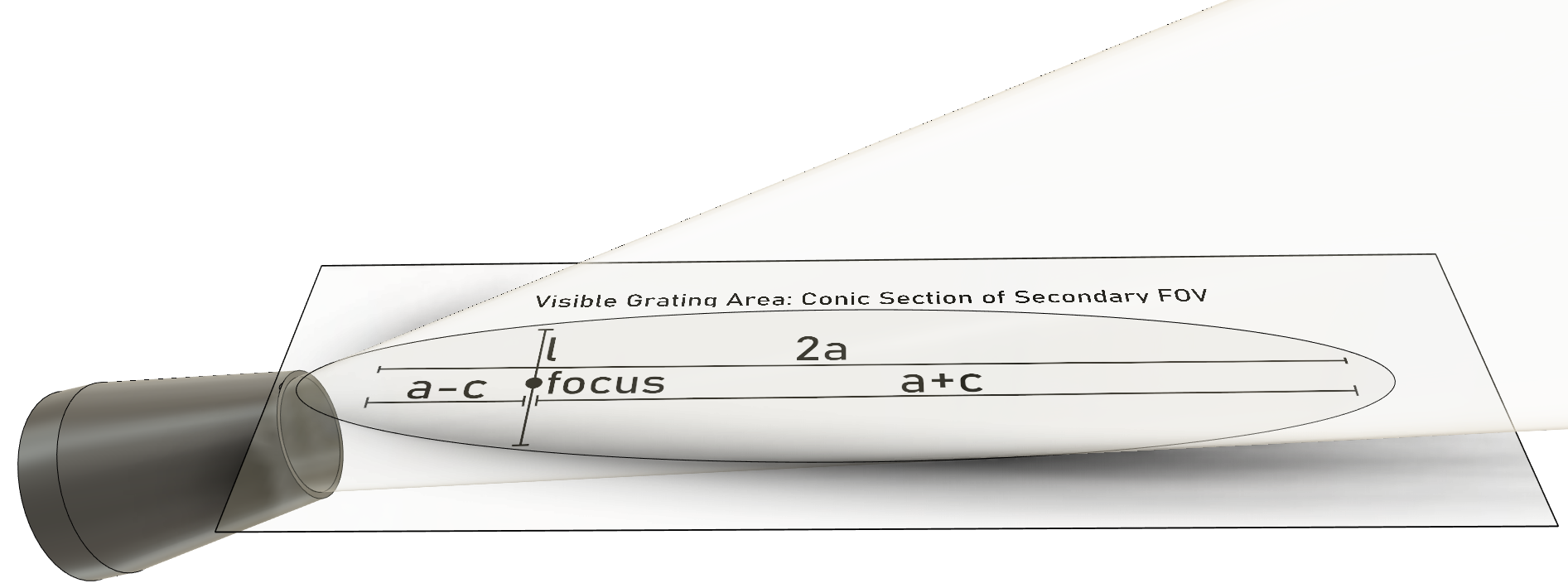}
\end{tabular}
\end{center}
\caption 
{ \label{fig:geo}
Visible grating area as a conic section of secondary focusing element FOV. The secondary focusing element field of view is exaggerated in order to demonstrate that total collection area is truly a conic section. It is seen from the figure that the major axis is determined solely from the secondary aperture diameter and field of view, and the angle at which the secondary aperture axis intersects the grating.} 
\end{figure}

To quantify the enhanced single-wavelength light collection area incurred from a Dittoscope application, we must consider the properties of the secondary focusing element collecting the output light at some central exodus angle. Figure ~\ref{fig:geo} imagines a wide FOV telescope or Schmidt camera observing a grating at high exodus angle, demonstrating that the total area visible to the secondary focusing element is the conic section subtended by its field of view.

 The analyses given here utilize circularly symmetric secondary optics. This choice is rather arbitrary, and is meant to model the secondary focusing element after the most conventional arrangements of telescopes/cameras. The grating is chosen to be rectangular in all figures as an aid to comprehension. In a practical application, the optics of the secondary might be rectangular in order to utilize the full area of the grating. Conversely, the grating area could be cut down to the elliptical grating area, or recorded/etched on an elliptical substrate.

The conic section area depicted in Figure~\ref{fig:geo} is the total area visible to the secondary focusing element. Though it will be seen later that each wavelength is collected from a smaller sub-area of the overall visible grating area, it is important to quantify this region in order to determine the necessary linear grating dimensions. One of the foci of the ellipse of visible grating area is identified by the intersection of the secondary focusing element optical axis with the grating surface. With one of the foci determined, geometrical properties of the visible grating area such as the linear eccentricity $c$, semi-major axis $a$, and semi-latus rectum $l$ can be derived from trigonometric calculations. The ellipse of visible grating area is determined by the semi-major axis and semi-latus rectum. These are given in terms of central exodus angle $\theta_E$ (the angle at which the secondary optical axis intersects the grating), secondary focusing element diameter $D$, and secondary field of view $\phi_{FOV}$:

\begin{equation}
\label{eq:conicarea}
a = \frac{D}{2}\left(\textrm{sin}(\pi/2-\theta_E) + \frac{\textrm{cos}(\pi/2-\theta_E)}{\textrm{tan}(\pi/2 - \theta_E - \phi_{FOV}/2)}\right)
\end{equation}

\begin{equation}
\label{eq:conicarea2}
l = \frac{D}{2}\left(1 + \frac{\textrm{tan}(\phi_{FOV}/2)}{\textrm{tan}(\pi/2 - \theta_E)}\right) \; .
\end{equation}

\noindent With these quantities in hand, we can determine the semi-minor axis $b$ and subsequently the total conic area $A$:

\begin{equation}
\label{eq:conicarea3}
b = \sqrt{(l*a)} \; , \; A = \pi*a*b \; .
\end{equation}

\begin{figure}
\begin{center}
\begin{tabular}{c}
\includegraphics[height=6cm]{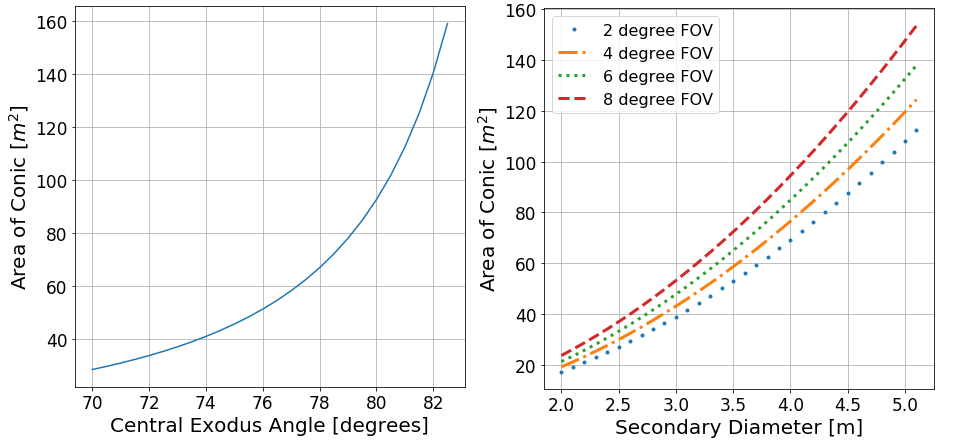}   
\\
(a) \hspace{5.1cm} (b)
\end{tabular}
\end{center}
\caption 
{ \label{fig:areas}
(a) Total grating area visible to the secondary focusing element, taken as the conic section subtended by its field of view. The secondary focusing element is assumed to have a diameter of 2.5m and FOV of 3$^\circ$, characteristics analogous to the Sloan Digital Sky Survey (SDSS) telescope. As the exodus angle approaches grazing at 90$^\circ$, the collection area is massively increased. (b) With the exodus angle constrained to 70$^\circ$, the secondary focusing element characteristics are varied. Wider secondary diameters and FOVs also result in massively increased light collection area on the grating.} 
\end{figure}

\begin{figure}
\begin{center}
\begin{tabular}{c}
\includegraphics[height=6cm]{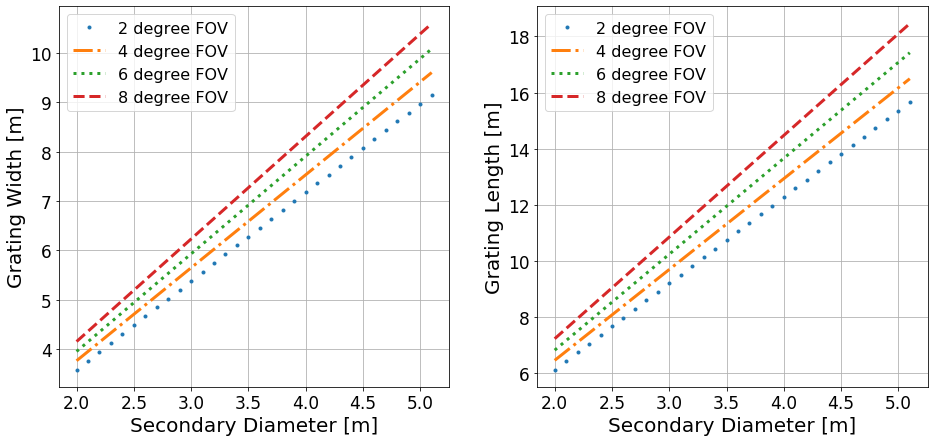}  
\\
(a) \hspace{5.1cm} (b)
\end{tabular}
\end{center}
\caption 
{ \label{fig:dimensions}
Linear grating dimensions (width (a) and length (b)) required to contain inscribed ellipse of visible grating area; shown as a function of the secondary focusing element diameter and field of view. In both panels of the figure, central exodus angle is constrained to 70 degrees.} 
\end{figure}

Figure~\ref{fig:areas} illustrates the scaling relationships of total conic area with the central exodus angle received by the secondary, as well as the diameter and FOV of the secondary focusing element. Figure~\ref{fig:dimensions} gives the required linear grating dimensions to inscribe (within a rectangular grating) the total visible conic area for a variety of secondary focusing element characteristics. In these figures the central exodus angle is constrained to 70$^{\circ}$.

 \subsection{Wavelength-Dependent Light Collection Area and Bandwidth}
\label{sect:area}

Though we have determined the total grating area visible to the secondary focusing element, the problem is far from being solved. While the secondary focusing element can receive photons from all points in the field of view, only light of certain wavelengths will diffract from the grating at the necessary angle to enter the aperture of the secondary. Furthermore, the angle required to enter the secondary aperture depends on location within the visible grating area.  It is for this reason that collection area must be treated wavelength by wavelength to get an accurate picture of the light collection area.

\begin{figure}
\begin{center}
\begin{tabular}{c}
\includegraphics[height=8cm]{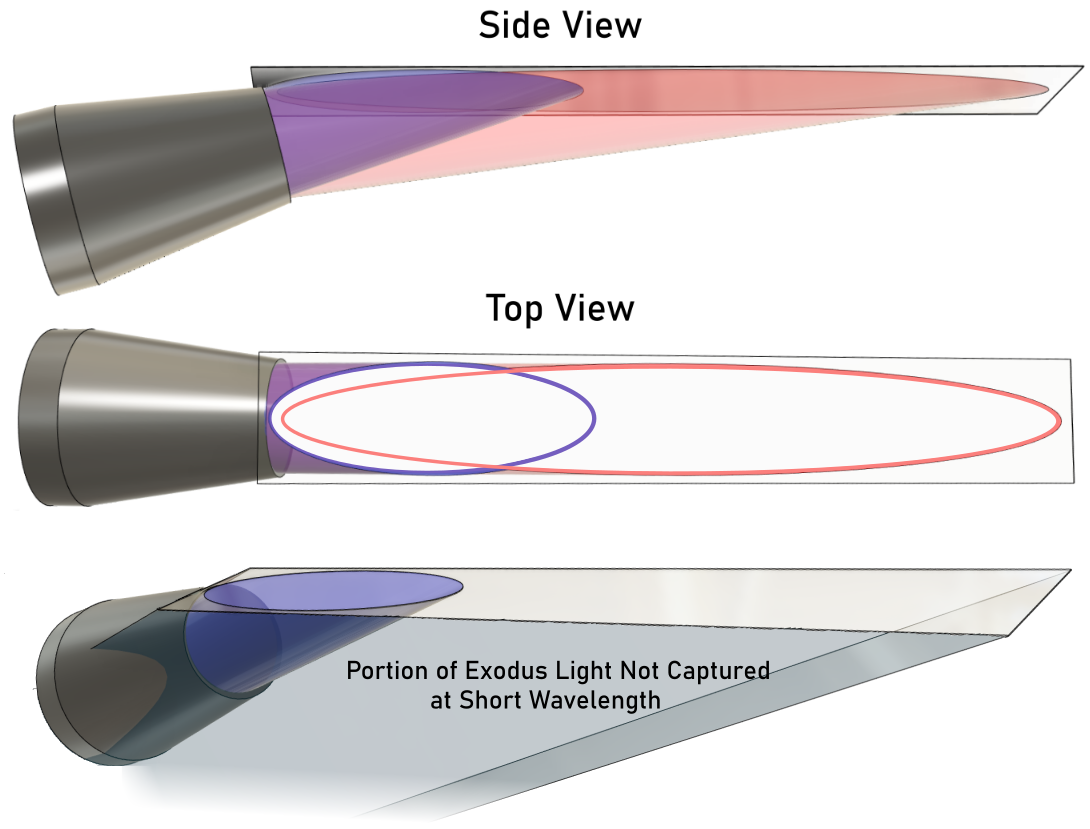}
\end{tabular}
\end{center}
\caption 
{ \label{fig:subareas}
The situation of Figure~\ref{fig:geo} is re-imagined in the case of multiple wavelengths, with the POG receiving light from a single distant point source. After diffraction from the POG, every wavelength emanates from the grating at a different angle, and visible grating area varies across the spectral bandwidth for any given object in the observatory FOV. As such, collection area must be averaged over the wavelength dependent `sub-areas' of the visible conic area. For any given wavelength, the collection area is the cylindrical section subtended by the secondary focusing element diameter. Shorter wavelengths (blue) are received at smaller exodus angles and therefore have smaller cylindrical-section collecting areas than longer wavelengths (red). At shorter wavelengths, a significant portion of the exodus light may not be captured by the aperture of the secondary focusing element (displayed in the bottom row). However, for situations with a smaller (and more realistic) secondary FOV, the portion of lost light at short wavelengths will be much less than the exaggerated situation of this figure.}
\end{figure}

\begin{figure}
\begin{center}
\begin{tabular}{c}
\includegraphics[height=6cm]{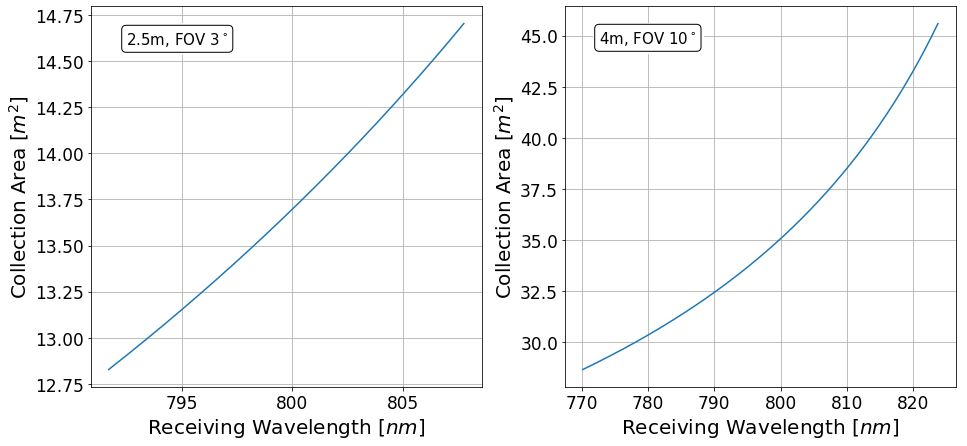}  
\\
(a) \hspace{5.1cm} (b)
\end{tabular}
\end{center}
\caption 
{ \label{fig:bandsub}
Collection area as a function of wavelength for the case of two different secondary focusing elements. Plot (a) imagines a secondary focusing element of diameter 2.5m and  FOV 3$^{\circ}$, while plot (b) has diameter 4m and FOV 10$^{\circ}$. Note that the FOV is specified because the limits of the FOV set the limits of the single object bandwidth. The diameters and fields of view are very large in both cases; this is intended to maximize the number of photons collected (this will be elaborated on later). For both telescopes the angle of incidence is 0$^{\circ}$, and the receiving/exodus angle is 70$^{\circ}$. We chose the grating spacing to ensure that the central wavelength would be 800\textrm{nm}; in this case the pitch is $\sim851$nm. Note that this figure gives the area of intersection of a cylinder with the grating ($A_{\lambda}$); the conic from a wide field of view would intersect a larger grating area, and collect light across the whole range of wavelengths. For an object at $\theta_{in}=0$, wavelengths outside of the plotted range would not intersect the secondary focusing element at an angle that is within its field of view.} 
\end{figure}

Light from an individual object within the observatory field of view is initially incident on the grating as a broadband coherent plane wave. After diffraction from the grating, the light is dispersed into many coherent plane waves exiting the grating in a wide range of angles at the corresponding wavelength (Eqn.~\ref{eq:gratingeq}). The secondary focusing element positioned at grazing exodus is only able to capture a narrow-band slice of these diffracted wavefronts, as only wavefronts diffracted within the angular range of the secondary focusing element FOV can enter the secondary aperture. The wavelengths included in this narrow (single-object) band vary with angle in the sky along the direction of dispersion. The end result is that, for each diffracted wavelength (from a single object within the observatory FOV), there will be a grating sub-area from which light enters the aperture of the secondary focusing optic. This sub-area of the total conic sectional area corresponds to a particular wavelength, and is determined from the cylindrical section subtended by the secondary focusing element aperture at the angle this light enters the secondary; this situation is illustrated in  Figure~\ref{fig:subareas}. Examples of light collecting area as a function of wavelength within the single-object bandwidth are given in Figure~\ref{fig:bandsub}.

Understanding the light collecting area for each object in the POG observatory FOV requires the application of a weighted average over these wavelength-dependent sub-areas, with weights corresponding to the spectral flux density of the observed object. In the interest of defining light collection power in all generality, we will assume a flat spectral flux density when computing the average collection area on the grating. This corresponds to conducting an unweighted average over all sub areas over all wavelengths in the band. Analytically, this is represented by an integral over the field of view. For a given ellipse of semi-major axis $a_{\lambda}$ and semi-minor axis $b_{\lambda}$, defined as a cylindrical section subtended by aperture diameter $D$ at an exodus angle of $\theta_E$. The average area over the single object bandwidth $\left< A_{\lambda} \right>$ may be found by integrating the cylindrical section over the valid range of angles $\theta_E - \phi_{FOV}/2$ to $\theta_E + \phi_{FOV}/2$:

\begin{equation}
\label{eq:area}
\left< A_{\lambda} \right> = \pi b_{\lambda} \left< a_{\lambda} \right> = \pi \left( \frac{D}{2} \right) \frac{D}{2\phi_{FOV}}  \int_{\theta_E - \frac{\phi_{FOV}}{2}}^{\theta_E + \frac{\phi_{FOV}}{2}} \frac{\;\mathrm{d}\theta}{\textrm{cos}(\theta)} = \frac{\pi D^2}{4\phi_{FOV}} ln\left(\frac{ tan \left(  \frac{\pi/2 + \theta_E + \phi_{FOV}/2}{2} \right) }{tan \left(  \frac{\pi/2 + \theta_E - \phi_{FOV}/2}{2} \right)} \right) \; .
\end{equation}

\noindent Each output angle in the valid angular range from $\theta_E - \phi_{FOV}/2$ to $\theta_E + \phi_{FOV}/2$ corresponds to a single wavelength in the single-object bandwidth. A weight parameter may be incorporated into the integral by relating the diffraction angle and corresponding wavelength to the spectral flux density of the object in question. The unweighted calculation was conducted for a variety of secondary focusing element characteristics. Results of these calculations are given in the second panel of Figure~\ref{fig:subareaswavs}.

It is worth noting that Equation \ref{eq:area} is formulated in the paraxial approximation of a diffraction grating. While a full analysis embodying non-paraxial diffraction theory\cite{Harvey:98} can be conducted, this is left for later publications. Equation \ref{eq:area} therefore fails to fully realize the effects of incident light in the direction perpendicular to dispersion (the oblique direction). However, in a non-paraxial analysis, deviation of wavelength-averaged collection area from that given in Equation \ref{eq:area} will be minor, since the cylindrical section sub-areas scale much more rapidly with angular changes in the diffraction direction as opposed to the oblique direction (due to the extreme tilt angle of the grating in the diffraction plane). This deviation will also be increasingly minor for smaller FOVs of the secondary focusing element, as the conic-sectional grating area becomes increasingly equal to the cylindrical section area as the secondary field of view decreases.

Given the impressive scaling of both the total conic sectional (Fig.~\ref{fig:areas}a) and averaged cylindrical collection areas (Fig.~\ref{fig:subareaswavs}b) with angle of exodus, one might assume that an arrangement with exodus angle close to 90 degrees will yield the greatest number of collected photons. Sadly this is not the case, as the extremely high dispersion near grazing exodus results in narrowing of the bandwidth available to the observatory for any single source in the FOV. The functional relationship, derived directly from the grating equation, reads:

\begin{equation}
\Delta \lambda = p(\textrm{sin}(\theta_{E} + \phi_{FOV}/2) - \textrm{sin}(\theta_{E} - \phi_{FOV}/2)).
\end{equation}

\noindent Application of this relationship to a sample of secondary focusing element characteristics and exodus angles yields Figure~\ref{fig:subareaswavs}a. Note that this result is independent of grating efficiency, which could also vary as a function of exodus angle. Typically one might expect efficiency to decrease with increasing exodus angle, but this relationship would depend on the details of the grating itself (e.g. blaze angle and groove profile).

\subsection{Wavelength-Dependent Spectral Resolvance}
\label{sect:spectdep}

In Section \ref{sect:area}, the collection area was seen to vary as a function of exodus angle/wavelength received by the secondary focusing element. Since the spectral resolvance of a grating in the first diffraction order is simply $R = N = L/p$ (the total number of illuminated fringes), the resolvance must also change as a function of exodus angle/wavelength. For a given exodus angle, the length of the illuminated portion of the grating is $L = D \textrm{sec}(\theta)$.  For a single object in the observatory FOV, the spectral resolvance will take the form:

\begin{equation}
\label{eq:resolver}
R = N = \frac{D}{p}\textrm{sec}(\theta) \; ,
\end{equation}

\noindent where $\theta$ is a specific exodus angle from the grating (within the range $\theta_E - \phi_{FOV}/2$ to $\theta_E + \phi_{FOV}/2$), D is the diameter of the secondary focusing element, and p is the grating pitch. Equation \ref{eq:resolver} was obtained by substituting $L = D \textrm{sec}(\theta)$, the major-axis of the cylindrical section subtended by the secondary aperture, into the resolvance equation.

In a more realistic scenario than the exaggerated example shown in Figure \ref{fig:subareas}, the secondary FOV will be small compared to the central exodus angle $\theta_E$. Because the wavelength range of each object observed is small, the resolvance ($R = \lambda/\Delta\lambda$) does not vary much across the bandwidth. Because the exodus angle is the same regardless of angle of incidence, the resolvance of each object is similar, even though objects observed at various angles of incidence would be observed at different wavelengths.

\subsection{Field of View}
\label{sect:fovy}

\begin{figure}
\begin{center}
\begin{tabular}{c}
\includegraphics[height=6cm]{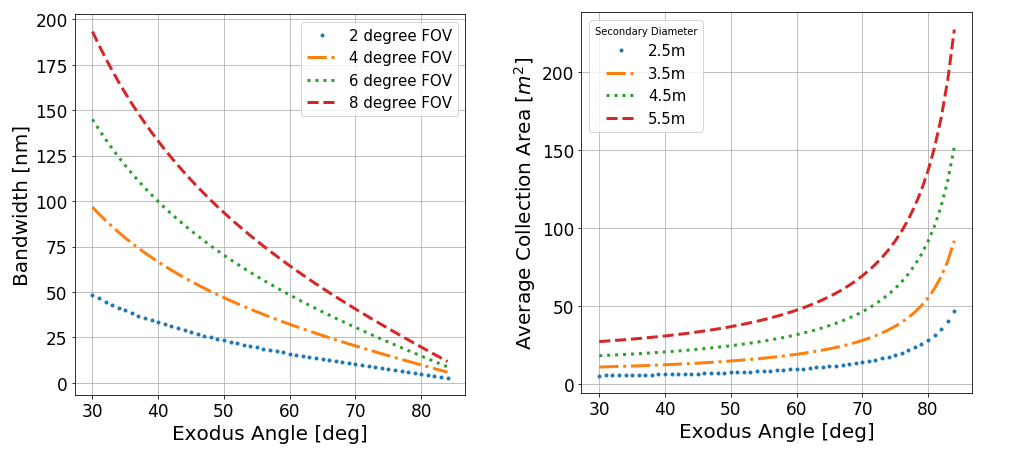}  
\\
(a) \hspace{6cm} (b)
\end{tabular}
\end{center}
\caption 
{ \label{fig:subareaswavs}
Plot (a): Single-object bandwidth as a function of exodus angle for a variety of secondary focusing optic fields of view. The telescope diameter is constrained to 2.5m. Note the negative scaling of single-object bandwidth with increasing exodus angle. Plot (b): Average collection area over the bandwidth, under the assumption of a flat spectral flux density, calculated as a function of central exodus angle for a variety of secondary focusing element apertures. The field of view is constrained to 3$^{\circ}$. The positive scaling of collection area with exodus angle to the secondary focusing element is quite apparent. Taking both plots into consideration, it is clear that the large light collection areas of plot (b) are mitigated by the narrow single-object bandwidths of plot (a).} 
\end{figure}

Although the addition of a POG to collect light does not significantly increase the number of photons collected, the high dispersion of the POG at grazing exodus does result in a massive increase to the field of view of the observatory. For any Dittoscope design composed of a secondary focusing optic/camera pointed at a plane grating, the observatory field of view along the axis of the grating is determined by the wavelength range at which observable sources emit, diffraction efficiency in the vicinity of the blaze angle (in the case of a blazed grating), and the free spectral range of the first diffraction order. Note that limitations of the free spectral range can potentially be overcome by the use of optical filtering and related techniques. The field of view perpendicular to the grating (oblique angles of incidence on the grating) is determined by the field of view of the secondary focusing element.

\begin{figure}
\begin{center}
\begin{tabular}{c}
\includegraphics[height=5.5cm]{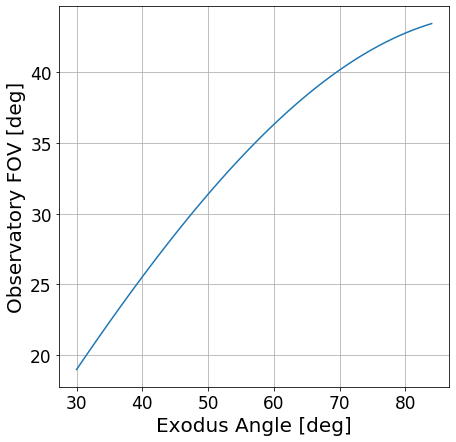}
\end{tabular}
\end{center}
\caption 
{ \label{fig:fov}
Observatory field of view in the grating direction as a function of exodus angle, for the assumption of 3$^{\circ}$ secondary FOV and 500$\textrm{nm}$ - 1100$\textrm{nm}$ bandwidth. It is seen that the field of view in the direction of the grating is massively increased compared to that of the secondary focusing optic in isolation. Many potential Dittoscope applications hinge on this massively increased field of view as a method to survey a very large patch of sky. Note that the observed single-object bandwidth will vary along the angle on the sky aligned with the long axis of the grating.} 
\end{figure}

Though the specific application  in question will ultimately determine the observatory FOV, an arbitrary wavelength range must be chosen in order to quantify the potential benefits of a POG design over conventional telescopes. If we assume a secondary focusing element with a field of view of 3 degrees, and a desired wavelength range of 500$\textrm{nm}$ - 1100$\textrm{nm}$, the overall observatory FOV is displayed in Fig~\ref{fig:fov} as a function of exodus angle. The 500$\textrm{nm}$ - 1100$\textrm{nm}$ range is chosen to have some overlap with the spectral range of conventional optical telescopes, but is consciously chosen to favor longer wavelengths, since large area gratings become increasingly difficult to manufacture with smaller groove-spacing (pitch).

Figure~\ref{fig:fov} demonstrates that high exodus angles result in a massively increased field of view relative to the secondary focusing element in isolation. This increase in field of view is among the most promising properties of the Dittoscope concept. Note, however, that the observed wavelength range for any individual source is quite narrow, and the detected wavelength varies over the entire detectable wavelength range, as a function of angular position in the sky along the long axis (dispersion direction) of the grating.

\subsection{A POG Figure of Merit: Modified \'etendue } 
\label{sect:etendue} 

 Among the most peculiar characteristics when assessing the merit of a hypothetical POG design is the rapid scaling of average collection area with exodus angle, coupled with the negative scaling of the per-object bandwidth (Fig~\ref{fig:subareaswavs}). Reduced bandwidth clearly leads to reduced photon flux, as a smaller slice of the spectral flux density is sampled for each object. This effect acts to offset the massive gains in light collecting area when moving to extreme exodus configurations.

This pair of counteracting scaling relationships inspires a new figure of merit similar to the widely used $\textrm{\'etendue}$ $=\Omega*A$, which quantifies the effectiveness of a conventional telescope in collecting photons. This is less useful for a POG observatory with a secondary focusing element at grazing exodus, since the \'etendue does not account for the severe restriction in the observable single-object bandwidth. Instead, we define a modified \'etendue:

\begin{equation}
\textrm{modified} \; \textrm{\'etendue} \; = \Delta \lambda * \Omega * \left< A_{\lambda} \right> \; .
\end{equation}

\begin{figure}
\begin{center}
\begin{tabular}{c}
\includegraphics[height=6.0cm]{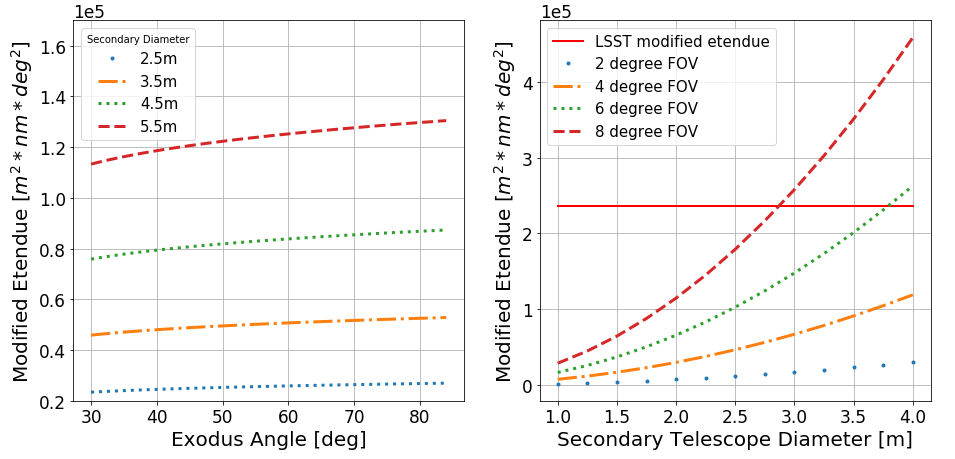}
\\
\hspace{1cm} (a) \hspace{6cm} (b)
\end{tabular}
\end{center}
\caption 
{ \label{fig:etendue}
Plot  (a): modified \'etendue as a function of exodus angle for a variety of secondary focusing optic diameters, a secondary focusing element FOV of 3 degrees is assumed. Note the meager scaling of modified \'etendue with exodus angle, resulting from the decreased single object bandwidth in conjuction with increasing FOV and light collection area. Plot (b): modified \'etendue as a function of secondary focusing optic diameter for a variety of secondary focusing optic FOVs, an exodus angle of 70 degrees is assumed. The modified \'etendue of LSST with no filter is given for comparison. Note the much more favorable scaling of modified \'etendue with respect to the parameters of plot (b) in comparison to plot (a). It is apparent that secondary focusing element characteristics may play a more important role in Dittoscope performance than the exodus angle for a given configuration.} 
\end{figure}

\noindent With this definition we calculate the modified \'etendue for a sample of POG characteristics. The results are given in Fig~\ref{fig:etendue}.

Figure~\ref{fig:etendue}a shows that modified \'etendue exhibits only modest scaling with increasing exodus angle, demonstrating that the massive gains to collection area shown in Figure~\ref{fig:subareaswavs}b are largely subdued by the negative bandwidth scaling of Figure~\ref{fig:subareaswavs}a. Rather than exodus angle, secondary focusing element characteristics of diameter and FOV emerge as the dominant factor in the scaling of modified \'etendue (Fig~\ref{fig:etendue}b). However, the grating collects a large amount of light in an extremely narrow band. This property may preserve the POG as a potentially useful component for specific science objectives (e.g. faint high-resolution spectral features at high angular resolution).

\section{Angular Resolution of POG Telescopes}
\label{sect:res}

\subsection{Diffraction Limit}

We show here that the angular diffraction limit is determined from the POG length, and not by the size of the secondary focusing element. This is because the angular distance between two objects is magnified by the grating.\cite{Ditto14}

Imagine two objects that are separated by a small amount $d\theta _{in}$ and observed at the same wavelength. By taking the derivative of the grating equation (Eqn.~\ref{eq:gratingeq}) at constant wavelength, we find that:

\begin{equation}
 \frac{d \theta_{out}}{d \theta_{in}} = \frac{\textrm{cos}(\theta_{in})}{\textrm{cos}(\theta_{out})} \;.
\end{equation}

\noindent In order to see this light on axis in the secondary focusing optic, $\textrm{cos}(\theta_{out})$ $\sim$ $D/L$, where $L$ is the length of the grating intercepted by the secondary focusing element and $D$ is the diameter of the telescope. For $\theta_{in} = 0^\circ$, $Dd\theta_{out} = Ld\theta_{in}$. The diffraction limit (in terms of $d\theta_{out}$) of the secondary focusing element is proportional to the diameter of the telescope. However, if we map the angle that is seen by the secondary focusing element to the sky ($d\theta_{in}$), that diffraction limit is proportional to the length of the grating. So in the limit of perfect gratings and no atmosphere, the Dittoscope can resolve objects that are aligned along the length of the grating at the resolution of $d\theta_{in} \sim \lambda/L$. The angular magnification of the POG is similar to the angular difference magnification of conventional telescopes, which comes about as a consequence of conservation of \'etendue. The Dittoscope is not diffraction limited by the secondary focusing element aperture, in the same manner that an ocular equipped telescope is not diffraction limited by the pupil of the human eye. 

\subsection{Atmospheric Seeing Simulations}
\label{sect:see}

\begin{figure}
\begin{center}
\begin{tabular}{c}
\includegraphics[height=10cm]{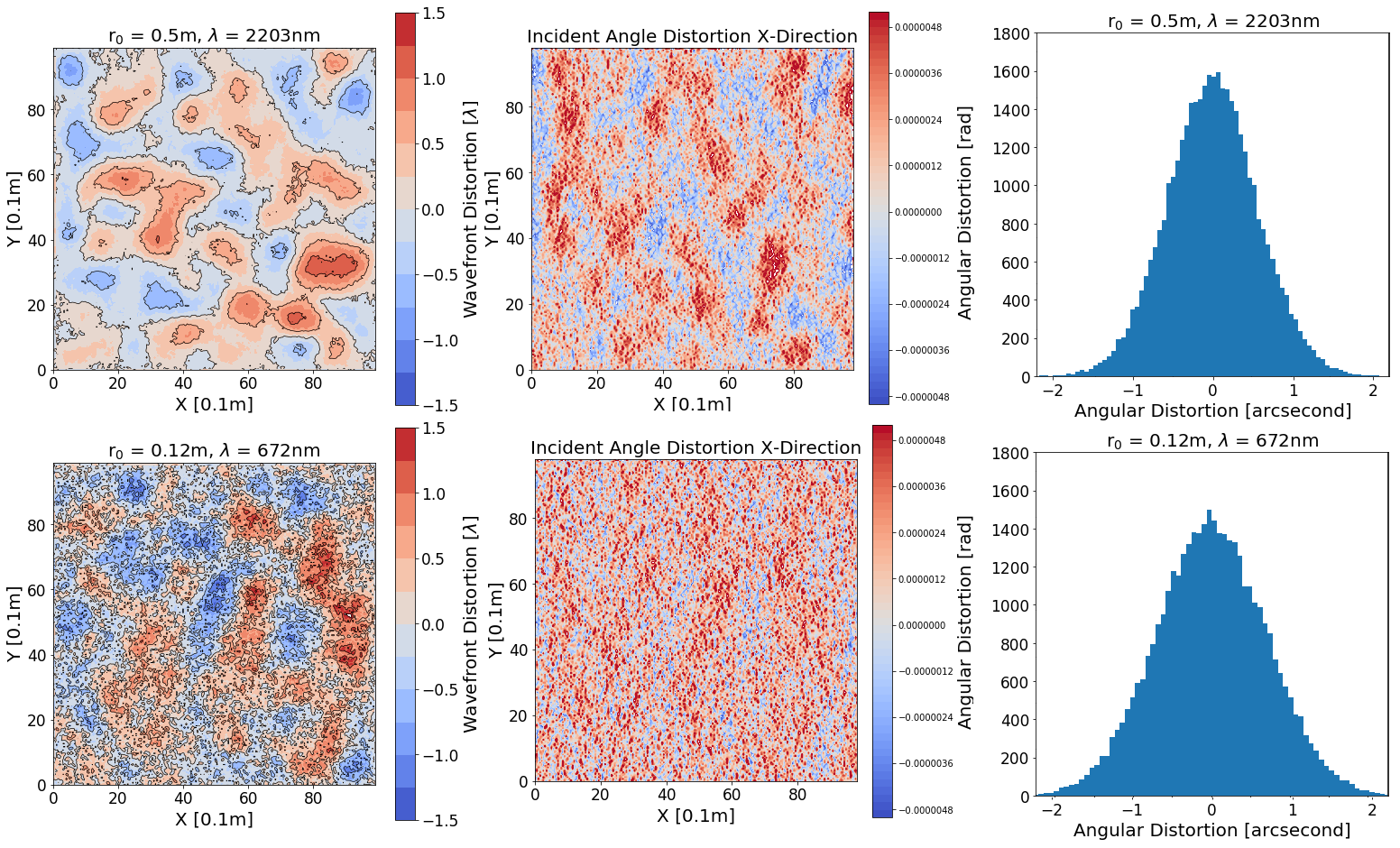}
\end{tabular}
\end{center}
\caption 
{ \label{fig:seeing}
Angular distortion for two simulations of differing coherence length under the same seeing conditions (coherence length is wavelength dependent). Theory dictates a coherence length of 0.5m and wavelength of 2203\textrm{nm} (top row) should result in an angular resolution of 0.88 arcseconds, and a coherence length of 0.12m and wavelength of 672\textrm{nm} (bottom row) should produce an angular resolution of 1.12 arcseconds. The rightmost plots demonstrate that the resulting angular distortion is in line with theoretical expectations.} 
\end{figure} 

For a ground-based Dittoscope, one must consider the effect of the atmosphere on the wavefront that reaches the POG. At sufficient distance, the radiation from a star will form a plane wave that is incident on the Earth. However, the atmosphere will cause wavefront distortions in the light that actually makes it to the Dittoscope.
Because the POG could potentially be tens or hundreds of meters long, it is much larger than the coherence length in the atmosphere. Therefore, the wavefront distortions will average out over the area of the POG, producing a blurring of the light from a point source that is similar to the `Airy disk' of a diffraction limited telescope (the time-averaged speckle distribution has the functional form of a diffraction limited Point Spread Function, PSF), but created instead from shifting diffraction patterns, or `speckles' that evolve on timescales of milliseconds.

To understand the effects of atmospheric seeing on the angular resolution of a ground based POG design, a simple simulation of atmospheric wavefront distortion was built to verify the seeing PSF predicted by theoretical considerations. This simulation generates sample wavefront distortions by using emergent phenomena to mimic the pseudo-random distributions of turbulent cells in the upper atmosphere. Sample data was generated on a $10\textrm{m} \times 10\textrm{m}$ grid with a grid spacing of $0.05\textrm{m}$. At initialization, random points of high wavefront distortion are injected into the grid, and allowed to spread by repeated application of Gauss-Seidel relaxation. At each relaxation step, a random fluctuation of tunable magnitude occurs at each grid point. By tuning the number of relaxation steps, degree of fluctuation at each point, and the number/intensity of weighted points in the initialization grid, phase distributions heuristically similar (coherence length at each point in the grid) to real phase-screen data were generated. The sky was modeled from a contour plot of phase-screen data taken over a $12\textrm{m} \times 12\textrm{m}$ area on Mauna Kea, Hawaii\cite{Monn}.

With the grid of simulated phase-screen data generated, sample light rays were produced by applying a triangular mesh to the grid. Taking the normal to each mesh face as an incoming light ray, and passing these sample rays through the grating equation. The resulting PSF in the focal plane can be visualized by a histogram of the angular distortions of each ray after passing through the grating equation.

Figure ~\ref{fig:seeing} shows incident angle distortions generated from simulation data. The histograms of incident-angle distortion are equivalent to the distortion incurred by a 1m x 100m collector (or any other 100m$^2$ primary grating), regardless of the fact that they are calculated from a 10m x 10m area; this is due to the isotropic nature of the wavefront distortions.

The seeing of, say, 1.5" becomes a blurring in the wavelength direction. In the case of our benchmark design, the wavelength blurring is $\sim$0.006\textrm{nm}. The spot size that this produces after light from the secondary focusing element focal plane is dispersed with a second grating depends on how effectively the second grating reconstructs the sky. In principle, if a second spectrograph is capable of identifying the wavelength and position of every photon that hits the secondary focusing element focal plane, that wavelength blurring will turn back into a blurring in angle that is equal to 1.5" in the final image. That is because if you detected the wavelength exactly, you would know to within 1.5" where the photon originated. This blurring precludes diffraction limited imaging, as it does in all ground-based observations.

\section{Disambiguating Objects in the Focal Plane}
\label{sect:focal}

We now turn our attention to the endpoint of the optical system, the arrangement of detectors in the focal plane of the secondary focusing element. Observation of a large area diffraction grating by a conventional telescope results in a situation synonymous with the familiar `grism-image', in that each object is spread out into a spectrum in the grating direction. The difference between the sort of image received with a grating observed at a grazing angle, as opposed to a low-dispersion grism, is that there will be no chosen `central wavelength' with which to disambiguate positions of objects within the FOV. Furthermore, in configurations approaching grazing exodus, a vanishingly small portion of the object spectrum is spread across the entire detector in the dispersion direction. Objects that are aligned with the grating will overlap over the entire length of the field of view, which could be tens of degrees or more (see Section~\ref{sect:fovy}).

The discussion of Section ~\ref{sect:fovy} reveals that the single-object bandwidth is determined by the secondary field of view, in combination with the relevant wavelength range of observation/emission. The definition of single object bandwidth as the wavelengths received at the margins of secondary FOV demands that the received light be spread across the entire detector plane defined by the secondary FOV. In this way it could be extremely difficult to distinguish the true positions on the sky, as objects seen in the focal plane will consist of many overlapping spectra. In theory, this issue can be resolved if the wavelength of received light is known at every position in the detector plane, as each wavelength received at a unique angle by the secondary can be mapped to a unique location on the sky. This approach presents its own difficulties; to achieve the optimal angular resolution, the spectrograph used to disambiguate focal plane wavelengths must be of similar resolving power to the POG.

Highly chromatic effects also plague the imaging capability of large area zone-plate concepts\cite{Eye,Moire}. In this case a holographic `corrector' is often proposed as a sort of anti-function to the aberration incurred by the primary\cite{patent,10.1117/12.7977006}. Unfortunately such an approach is likely impossible for POG designs, owing to the fact that there is no optical component that can act as an effective anti-function to the chromatic distortion introduced by the POG. 

\subsection{Dual-Dispersion}

In THE MOST, a slit is placed at the focal plane of the secondary focusing element, oriented parallel to the grooves in the grating. Light from this slit is conveyed onto a secondary dispersing element. This technique, coined as dual dispersion, ensures that every location on the detector plane corresponds to a unique wavelength and position on the sky.  The secondary dispersing element is shown conceptually in Figure \ref{fig:pog} as a prism in combination with an imaging camera. In practice, the secondary dispersing element would likely take the form of a conventional multiple-input spectrograph placed in the focal plane.

To measure all photons captured by the secondary focusing element (a condition assumed in all previous calculations), each pixel in the focal plane of the secondary must act as a slit or optical fiber, feeding into some sort of spectrographic instrument. The wavelength and direction from which each photon originated are disambiguated by dual dispersion. While this technique is an elegant method for disambiguation of all photons received at the detection plane, it results in a colossal increase in the number of necessary pixels in the detector. In a normal telescope application, the number of pixels is set by the resolution and field of view, in a dual dispersion architecture, the number of pixels is compounded with the wavelength resolution and observation bandwidth:

\begin{equation}
\label{eq:pixels}
\textrm{\#} \; \textrm{of} \; \textrm{pixels} = \frac{\pi}{4} \left( \frac{\textrm{secondary}_{FOV}}{\textrm{secondary}_{\textrm{resolution}}} \right)^2   *  \left( \frac{\textrm{bandwidth}}{\lambda_{\textrm{ resolution}}} \right).
\end{equation}

\noindent The first two terms of Equation~\ref{eq:pixels} describe the number of pixels required for a conventional telescope (the secondary focusing element) to take full advantage of its diffraction limited angular resolution and FOV. The last term in Equation~\ref{eq:pixels} reflects the number of resolution elements required for the spectrograph to disambiguate the multiple wavelengths received from a wide swath of the sky in the direction of dispersion. 

Drawing upon Figures~\ref{fig:dimensions},~\ref{fig:bandsub}, \&~\ref{fig:fov}, we can imagine a telescope consisting of a 2.5m secondary focusing element with a 3$^\circ$ FOV, viewing an 8m long grating at an exodus angle of 70$^\circ$. If we set the observable wavelength range to 500nm-1100nm (overall telescope bandwidth of 600nm), the grating pitch is constrained to 851nm by the grating equation. For this case, the field of view in the dispersion direction is 40$^\circ$ (again determined from the grating equation), leading to an overall FOV of 40$^\circ  \times \;$3$^\circ$. The resolvance of a grating in the first diffraction order is equal to the number of diffraction fringes. For the case of an 8m grating with 851nm pitch, we arrive at a resolvance of $\sim9.4\times10^6$, and a minimum resolvable wavelength difference of $\sim5.3\times10^{-5}$nm. The diffraction limited angular resolution of the secondary focusing element, which corresponds to the observatory resolution in the non-dispersing direction, is approximately $1.4\times10^{-5}$ degrees (this is the angular resolution at the high frequency side (500nm) of the observation bandwidth). Taking these quantities into account, we arrive at:

\begin{equation}
\textrm{\#} \; \textrm{of} \; \textrm{pixels} = \frac{\pi}{4} \left( \frac{3^\circ}{{1.4\times10^{-5}}^\circ} \right)^2  *   \left( \frac{600\textrm{nm}}{5.3\times10^{-5}\textrm{nm}} \right) \approx 10^{17} \; .
\end{equation}

This is undoubtedly an unreasonable quantity of pixels, especially considering that in a dual-dispersion architecture, the overwhelming majority of these pixels will capture no photons. This calculation is performed under the assumption that all photons incident on the focal plane of the secondary are successfully disambiguated by dual dispersion. In this view, each pixel in the focal plane of the secondary represents a single `slit', or spectrographic fiber, that is subsequently fed into some sort of extremely high resolution spectrograph. 

This calculation assumes the full angular resolution of the POG and the secondary focusing element is achieved. While this situation is likely impossible, it serves as an upper bound to the number of pixels required. It is worth noting that a particular Dittoscope application may not require the full resolution limits of the observatory. Limiting the resolution could considerably reduce the number of pixels, provided the science objectives in question do not require the full spectral/angular resolution available. Note, however, that the resolution in wavelength and angular sky position in the dispersion direction are tied together, so one cannot simply choose to lower their resolution independently. Also, some imperfections in the grating will impact the resolution in angular position, most notably errors in groove quality and pitch regularity.

In the previous example, the upper-limit on the number of required pixels was informed by the physical size of the POG and secondary under the assumption of ideal angular resolution. If, instead, the science objectives inform the required resolution, we can construct an application in which dual dispersion requires fewer pixels to disambiguate incident photons. One example is the POG being used for its large field of view rather than high resolution (Section \ref{sect:HFOV}). If a given science objective only required an angular resolution of  $0.1^{\circ}$, then (assuming the same arrangement as the previous example) the approximately 40$^\circ  \times \;$3$^\circ$ swathe of sky would require $\sim 12,000$ pixels to represent each area of the sky (it is worth noting that this situation requires asymmetrical slits/fibers in the focal plane of the secondary focusing element). Plugging a $0.1^{\circ}$ difference into the grating equation yields a required spectrograph resolution of $\sim1.5\textrm{nm}$, and the number of pixels in each spectrograph channel would be $\sim 400$ based on the 600nm bandwidth of this arrangement. The number of required pixels for this hypothetical arrangement is approximately 5 million, which is easily achieved with current detector technology.

\section{Discussion: Reaping the Benefits of POG Designs}
\label{sect:pros}

We have shown that POG observatories present uniquely challenging properties. Massive increases in field of view and collection area (relative to the secondary focusing element in isolation) are coupled with a decreased per-object bandwidth and difficulty in disambiguating photons in the focal plane. The results of Section~\ref{sect:etendue}, namely the plot of modified \'etendue in Fig~\ref{fig:etendue}, demonstrate that only meager scaling of this figure of merit is achieved by approaching a grazing exodus configuration. Our results suggest that if the only goal is to collect as many photons as possible (that are undifferentiated by wavelength), it is simpler and more cost effective to use a conventional telescope rather than a POG observatory.

Given this distinction, is it possible to design a POG observatory that is competitive with conventional technology? Novel POG architectures may be useful for a variety of highly specialized science objectives. Any given Dittoscope design is not likely to be a generalist instrument, but for applications requiring extremely large fields of view, high angular resolution, multiple-object high resolution spectroscopy, or low aerial mass, a POG design may prove extremely competitive. 

\subsection{High Angular Resolution} 

In Section~\ref{sect:res}, it was shown that POG telescopes possess angular resolution equal to that of a conventional telescope with diameter equal to the grating length. For a space-based POG observatory, this presents a unique low-mass solution to observations requiring extremely high angular resolution (e.g. extrasolar planet observations, as described in a proceedings publication on the Diffractive Interfero Coronagraph Exoplanet Resolver, DICER\cite{10.1117/12.2629487}). This extremely precise angular resolution is only realized in one dimension, and only for a small bandwidth (typically on the order of nanometers). Any science objectives seeking to utilize this one-dimensional resolution will require POG architectures and observational strategies designed to work around these limitations.

\subsection{High Field of View} 
 \label{sect:HFOV}

If attention is shifted away from angular resolution in favor of the massive increase to the field of view outlined in Section~\ref{sect:fovy}, a ground based observatory becomes an intriguing possibility. The natural application for a high FOV low single-object bandwidth observatory is situational awareness (i.e. detection of Near Earth Objects, NEOs, as described in a proceedings publication on Trip Wire Optics, TWO\cite{Newberg20}). One challenge encountered with this application is disambiguation of focal plane objects with a reasonable number of detector pixels, a problem which has not been solved as of the time of this publication. Another challenge is the limit on angular resolution imposed by seeing without the aid of adaptive optics, as discussed in Section \ref{sect:see}.

\subsection{High Spectral Resolution: Radial Velocity Measurements} 

All POG observatories will typically feature extremely high spectral resolution, with this resolution being proportional to the grating length. This extremely high spectral resolution is generally taken in a vanishingly small bandwidth for any single source, requiring highly specific and narrow spectral features to be a useful property.

A natural application is that of radial velocity measurements of stellar spectral lines. This approach may allow a Dittoscope to discover new exoplanets by the radial velocity method, and measure exoplanet masses for those that were discovered by their transits. If high enough spectral resolution can be achieved, a Dittoscope could possibly be used to map the accelerations of stars in the Milky Way. If the accelerations of stars within the Milky Way were known at the level of cm/s, new limits could be placed on the distribution of dark matter in the Galactic halo. In addition to technical feasibility of high resolution, we would also need to address issues of intrinsic line width and stability in astronomical objects, which is usually addressed by simultaneous observation of multiple lines.

\subsection{Rotating POG Observatories: Spectrographic Surveys} 

The narrow bandwidth properties of many POG arrangements could in theory be solved by operating the telescope in a mode analogous to a `drift scan'. By allowing the POG observatory to slowly rotate by means of orbital motion, the Earth's rotation, or by manual adjustment, the full spectrum of all objects could be built up over the course of a full transit. In a conventional telescope, all objects in a small field of view are captured with a large bandwidth, a drift scan serving to image more objects. A POG observatory takes a different approach, with many objects in a wide field of view captured in a sequence of narrow high-resolution bands, the drift scan serving to build up a wider spectrum for each object. Using such a technique, a spectral survey could be accomplished with little or no active pointing of the observatory.

\subsection{Future Considerations}
\label{sect:issue}

In these analyses, many important features and requirements of POG technology have not been fully addressed. Among the most important unaddressed considerations are:

\begin{itemize}
\item Optical efficiency of diffraction gratings at or near grazing exodus: \\ Many efficiency characteristics depend upon a variety of factors including the wavelength-range in consideration, angle of exodus, blaze angle, optical thin-films/coatings, and whether the grating is holographic or surface relief.
\item Cost effectiveness of large area diffraction gratings: \\ Current state-of-the-art methods\cite{Xu} can produce gratings of area $\sim1\textrm{m}^2$ via Reactive Ion Etching (RIE) techniques. However, these methods are currently incapable of producing larger gratings. New methods of manufacture (i.e. large area holography or RIE) must be developed for many Dittoscope concepts to become viable.
\item Optical characteristics: \\ Surface figure tolerance must be characterized. The effective Airy PSF after focusing from a large area grating remains to be fully understood. Laboratory experiments must be devised to address these concerns. Issues of efficiency at grazing, line spacing tolerance, and groove profile/blazing have yet to be fully explored for this arrangement.

\item Detectors: \\ Many Dittoscope architectures may require light detection at the single-photon level. Many emerging technologies such as the Skipper CCD\cite{PhysRevLett.119.131802} and Transition Edge Sensor (TES) bolometer\cite{Nagler} may enable POG technology in the coming years.
\end{itemize}

\noindent In order for POG technology to be fully realized as a feasible concept, all of the above requirements must be addressed in future works. 

\section{Conclusion} 

The basic telescope design consists of a long flat POG that diffracts light into a secondary focusing element of specified diameter and field of view. In the focal plane of the secondary optic, a small wavelength range of each observed object is spread out in the dispersion direction of the grating. The position of the spectrum depends on the angular sky position in the oblique direction perpendicular to the dispersion direction of the grating. The wavelength range depends on the angular sky position in the dispersion direction of the grating. Position in the sky in the dispersion direction can be determined if the wavelengths are separated by an additional disperser.

In addition to considerations for specific applications, we have approached the analysis of non-focusing POG telescopes from a broad and general standpoint, seeking to find basic metrics for comparing POG technology to more conventional astronomical observatories. In our analyses, we have attempted to make as few assumptions as possible regarding the components of the optical system, and set constraints on the secondary focusing element only when strictly necessary. Using this approach, we have derived many simple relationships connecting a Dittoscope's most basic physical parameters to useful metrics quantifying its ability to collect astronomical photons:

\begin{itemize}
\item In principle, dispersion of POGs observed at grazing exodus is extremely high; $R$ is formally in the millions for grating lengths in excess of several meters. Extremely high resolution spectra could be contemplated.
\item The use of a non-focusing POG at grazing exodus for optical leverage drastically increases the light collection area over that of the secondary focusing element alone. However, this large area collects photons in a small bandwidth for each object, and the observed wavelength range varies across the sky.
\item Evaluation of the wavelength-dependent light collection area, observatory field of view, and single object bandwidth led to the definition of a modified \'etendue $\left( \Delta \lambda * \Omega * \left< A_{\lambda} \right> \right)$; a useful metric for evaluating the light collection efficiency of a Dittoscope. 
\item The modified \'etendue scales with secondary focusing element characteristics of diameter and field of view, rather than the properties or position of the POG itself. This result is surprising considering the massive increase in collection area achieved at grazing exodus. However, the increase in collection area is effectively subverted by narrowing of the single-object bandwidth.
\item Dittoscopes exhibit a massively increased field of view compared to the secondary focusing element in isolation. A field of view in excess of $40^{\circ}$ is theoretically possible. Note, however, that this FOV is observed in different wavelength ranges in the dispersion direction of the grating.
\item The angular resolution (in the direction of dispersion) of a POG at grazing exodus is proportional to the grating length. The proportionality relationship is conceptually similar to the angular enhancement properties of a conventional telescope from conservation of \'etendue. The diffraction limit in the dispersion direction is $\sim\lambda/L$. This could make a Dittoscope an attractive option for applications requiring extremely high resolution.
\item Atmospheric seeing has similar effects on a ground-based POG telescope as it does on a conventional ground-based telescope. In order to capitalize on the full angular and spectral resolution available to the POG, a Dittoscope must be placed in space.
\item Due to the multiple overlapping spectra in the focal plane of the secondary focusing element, disambiguation of focal plane objects is a major problem for theoretical POG applications. Dual dispersion is currently the only known method for disambiguating all photons received at the focal plane. However, choices must be made to avoid requiring an
unreasonable number of pixels in the detector.

\end{itemize}

Given these findings, we suggest a few circumstances in which non-focusing POG telescopes might be advantageous. The potential for low aerial mass deployment of POG observatories in space might enable cost competitive missions that are impossible to achieve using a conventional telescope. Applications necessitating extremely high spectral resolution, wide field of view, low aerial mass, or extremely high angular resolution, are well suited to benefit from an application of POG technology.

\appendix

\subsection* {Acknowledgments}
This work was
supported by NASA Innovative Advanced Concepts (NIAC) Phase II grant \\
80NSSC19K0973, Manit Limlamai, and a fellowship from the NASA/NY Space
Grant.


\bibliography{report}   
\bibliographystyle{spiejour}   


\vspace{2ex}\noindent\textbf{Leaf Swordy}  is a doctoral student at Rensselaer Polytechnic Institute. He received his BS and MS degrees in physics from Rensselaer Polytechnic Institute in 2018 and 2019 respectively. With a background in high-purity liquid noble gas systems and general R\&D for the XENON1T and nEXO experiments, his research now centers on theoretical telescope architectures utilizing holographic primaries.

\vspace{2ex}\noindent\textbf{Heidi Jo Newberg}
 is a Professor in the Department of Physics, Applied
Physics, and Astronomy at Rensselaer Polytechnic Institute. She made
significant contributions to the Supernova Cosmology Project, the Sloan
Digital Sky Survey, the Chinese LAMOST project, and currently runs the
MilkyWay@home volunteer supercomputer. She is a Fellow of the American
Physical Society and has published in diverse areas of astrophysics
including discovery of tidal streams and substructure in the Milky Way
halo, discovery of disequilibrium in the Milky Way disk, properties of
stars, astronomical surveys, supernova searches, discovery of dark
energy, the constraining the spatial distribution of dark matter, and now Dittoscopes.

\vspace{2ex}\noindent\textbf{Thomas D. Ditto}
is an inventor and artist who proposed the “Dittoscope” in 2002. It derived from a
microscope of similar design which used grazing incidence as leverage. By reversing the angles of
input and output, the leverage shifted from microscopic examination to astronomical. Subsequently
the microscope was developed under research support from the NSF, and the telescope has been
studied under a Fellowship from NIAC.

\vspace{1ex}

\listoffigures

\end{document}